\documentclass[11pt]{article}

\input epsf.sty

\usepackage{graphicx,amsmath,amssymb}
\def\lromn#1{\uppercase\expandafter{\romannumeral#1}}

\begin{document}
\begin{flushright}
%\today \\
%${\cal M. Y.}$
\end{flushright}

\begin{center}
\begin{Large}
\textbf{
Persistent magnetization  at neutrino pair emission
}
\end{Large}

\vspace{1cm}

 M. Yoshimura

Research Institute for Interdisciplinary Science,
Okayama University \\
Tsushima-naka 3-1-1 Kita-ku Okayama
700-8530 Japan

\vspace{7cm}

{\bf ABSTRACT}

\end{center}

Measurement of parity violating magnetization  is proposed as
a means to determine neutrino properties such as
Majorana/Dirac distinction and absolute neutrino masses.
The process we use is radiative neutrino pair emission
from a collective and coherent body of lanthanoid ions doped in host crystals. 
A vector-component of electron spin flip 
parallel to  photon direction emitted from ion excited state generates 
this type of magnetization
which is stored for a long time in crystals till spin relaxation time.

\vspace{4cm}

Keywords
\hspace{0.5cm} 
Majorana neutrino,
neutrino mass,
parity violation,
neutrino pair emission,
magnetization of  lanthanoid ions,
SQUID

\newpage

\section
 {\bf Introduction}

Measurement of parity odd quantity in radiative neutrino pair emission
(RENP) \cite{renp overview} is of great use to distinguish
the process involving weak interaction from purely
QED processes \cite{pv ysu}.
An interesting quantity of this nature
is magnetization parallel to the emitted photon direction along with neutrino pair emission:
expectation value in the final state $| f\rangle $,
$2 \mu_{B} \langle f|     \hat{ k}\cdot \vec{S}| f \rangle $,
where $\hat k$ is the unit vector along the emitted photon momentum
and $\vec{S}$ is the electron spin operator with $\mu_B$ the Bohr magneton.
This quantity is parity  odd and time reversal (T) even.
The generated magnetization persists till spin relaxation time.

Parity odd quantity may emerge from interference term of parity even and odd operators
in the weak hamiltonian $H_W$ of neutrino pair emission.
Writing the neutrino pair emission hamiltonian in terms of
the neutrino mixing matrix elements, $U_{ei}\,, i=1,2,3\,,
\nu_e = \sum_i U_{ei} \nu_i $,
the hamiltonian is given by
\begin{eqnarray}
&& 
H_W =
\frac{G_F}{\sqrt{2}}\, \Sigma_{ij}\,
\bar{\nu}_i \gamma^{\alpha}  (1 - \gamma_5)\nu_j \,
\bar{e}\left( \gamma_{\alpha} c_{ij} - \gamma_{\alpha} \gamma_5 b_{ij}
\right) e
\equiv
\frac{G_F}{\sqrt{2}}\left( {\cal N}_c ^{\alpha}
\bar{e} \gamma_{\alpha} e  - {\cal N}_c ^{\alpha} 
\bar{e}\gamma_{\alpha} \gamma_5 e
\right)
\,,
\label{weak cc}
\\ &&
b_{ij} = U_{ei}^* U_{ej} - \frac{1}{2}\delta_{ij}
\,, \hspace{0.5cm}
c_{ij} = U_{ei}^* U_{ej} - \frac{1}{2} ( 1- 4 \sin^2 \theta_w) \delta_{ij}
= b_{ij} + 2 \sin^2 \theta_w \delta_{ij}
\,,
\end{eqnarray}
where $\nu_i$ denotes a neutrino mass
eigenstate of mass $m_i$.
The dominant interference term of parity violation arises from product of spatial part of axial vector current 
of electron, 
the spin operator $\bar{e} \vec{\gamma}\gamma_5 e \sim \langle \vec{S}
\rangle $
(in the non-relativistic limit), and spatial part of vector current, the velocity operator 
$\bar{e} \vec{\gamma} e \sim \langle  \vec{v} \rangle = \langle \vec{p}/m_e \rangle$.
Roughly, magnetization caused by $(ij)$ neutrino-pair emission is proportional to
$\Re (b_{ij} c_{ij})$, assuming CP conservation, while RENP event rate
is proportional to $\Re (b_{ij}^2)$.
Experimentally, the weak mixing angle is given by
$1 - 4\sin^2 \theta_W \sim 0.046 \pm 0.00 64$.
Matrix elements of spin $\langle \vec{S} \rangle $
and velocity operator $\langle \vec{v} \rangle $ are usually of order unity 
vs a number less than $10^{-3}$, hence
interference terms are at least smaller by $10^{-3}$ 
than rate, a parity even quantity.
We shall quantify this ratio for lanthanoid ion we use as target.

Using an effective hamiltonian $H_W'$ of RENP, 
one calculates generated magnetization,
\begin{eqnarray}
&&
\sum_{\nu} \langle f | \hat{k} \cdot \vec{S} | f \rangle
| \langle f | H_W' | e\rangle |^2
= \sum_{\nu}   \langle e | H_W' | f \rangle
\langle f | \hat{k} \cdot \vec{S} | f \rangle 
\langle f | H_W' | e\rangle
\,,
\label {mag matrix el}
\end{eqnarray}
where neutrino variables, their helicities and momenta, are summed over.
Hamiltonian $H_W$, hence $ H_W'$ as well,
 contains both parity even and parity odd operators due to
weak interaction of neutrino pair emission.
One of hamiltonian matrix elements in eq.(\ref{mag matrix el})
 must be parity even and the other
parity odd, requiring interference of these.
Furthermore, 
in order to have a non-vanishing  
$ \langle f | \hat{k} \cdot \vec{S} | f \rangle $,
the state $| f \rangle$, an energy eigenstate of
unperturbed parity even hamiltonian $H_0$, 
must contain both parity even and odd components.
This peculiar situation, which never occurs in isolated atoms, arises
when ions are placed in crystals,
since crystal field acting on target ions provides parity mixing.
In other words, host crystals provide environmental parity violation.
Lanthanoid ions of 4f$^n$ ($n=11$ for Er$^{3+}$) 
system gives rise to 
what is called forced electric dipole,
as pointed out in \cite{van vleck} and its calculation method formulated in \cite{judd-ofelt}.
This is why we adopt lanthanoid ions doped in host crystals as targets.
Moreover, 4f electrons in lanthanoid ions are
insensitive to environment of host crystals due to
filled 5s and 5p electrons in outer shells, which give
large spin relaxation time, very important to magnetization measurement.
Lanthanoid ions doped with low concentration exhibit
paramagnetic property at room temperature.

There have been two proposed methods that use a collective and coherent body
of atoms to study still unknown neutrino properties \cite{renp overview},
\cite{ranp}.
The method we propose here is different from previous proposals
in that we detect accumulated effect remaining in target medium rather than
measuring individual events themselves.
If the accumulated magnetization is above the sensitivity level of detectors, for instance, 
a high quality SQUID, then 
the method is found to be very sensitive to Majorana/Dirac distinction,
as shown below.
Our approach does not assume the nature of neutrino masses,
but lanthanoid experiments can determine  whether neutrinos are
of Majorana or of Dirac type by measuring angular distribution of
magnetization caused by interference terms
intrinsic to Majorana  neutrino, but absent in the Dirac neutrino case
\cite{my-07}.
Measurement of smallest neutrino mass at several meV level in three flavor scheme
requires high statistics data, or use of
smaller Stark level transitions in crystals.
Since there exist a rich variety of J-manifold levels in some lanthanoid ions,
both measurement of Majorana/Dirac distinction and smallest neutrino mass
seems possible. 

Process without photon emission $|e \rangle \rightarrow
| g \rangle + \nu\bar{\nu}$ might be considered for
magnetization measurement, but estimate  gives
rate much smaller than detection level.
Radiative neutrino pair emission $|e \rangle \rightarrow
| g \rangle + \gamma+  \nu\bar{\nu}$ \cite{renp overview}
gives 10 orders of magnitude larger rates than the process without photon emission.
Raman stimulated radiative emission $\gamma + |e \rangle \rightarrow
| g \rangle + \gamma+  \nu\bar{\nu}$
\cite{ranp} is also conceivable \cite{ect},
but the present scheme is simpler and has a better sensitivity
to neutrino properties.

We use the natural unit of $\hbar = c = 1$ throughout the present work
 unless otherwise stated.

\section
{\bf Magnetization generated at RENP (Radiative Emission of Neutrino Pair)}

RENP de-excitation path of ion states is defined as
$| e \rangle \rightarrow | p \rangle \rightarrow | g \rangle $
with energies $\epsilon_e > \epsilon_p > \epsilon_g$.
This is a cascade process in which neutrino pair is emitted
at $| e \rangle \rightarrow | p \rangle$ followed
by stimulated photon emission at $ | p \rangle \rightarrow | g \rangle $:
$| e \rangle \rightarrow | g \rangle + \gamma_0 + \nu{\bar \nu}  $.

Probability amplitude of RENP including parity violating terms
is given by
\begin{eqnarray}
&&
{\cal M} = \frac{G_F}{\sqrt{2}} \frac{\vec{\mu}_{pg}\cdot \vec{B}_0 }
{\omega_0 - \epsilon_{pg} + i \gamma_p/2 }
 (\vec{S}_{ep} \cdot \vec{{\cal N}}_b+ \frac{\vec{p}_{ep}}{m_e}\cdot 
\vec{{\cal N}}_c  )
\,,
\end{eqnarray}
where $\vec{{\cal N}}_{b,c}$ is the neutrino pair emission current defined by eq.(\ref{weak cc}).
Magnetic dipole transition given by $\vec{\mu}_{pg}\cdot \vec{B}_0 $ in this formula is
usually dominant  for 4f$^n$ lanthanoid optical transitions among ion lower levels, where
$\vec{B}_0$  is RENP trigger field strength with its frequency given by $\omega_0$.
There may be a forced electric dipole transition of the form, 
$\vec{e}_{pg}\cdot \vec{E}_0 $, as well in lanthanoid ion transitions.
To maximize RENP rates, we irradiate the trigger beam at resonant frequency, 
$\omega_0 = \epsilon_{pg}$, equated to a level energy spacing.
Convolution integral of squared resonance function 
$\propto 1/\left( (\omega_0 - \epsilon_{pg})^2 + \gamma_p^2/4 \right)$
with the Lorentzian spectrum  gives $4/ (\Delta \omega_0 \gamma_p)$, hence
we use, in rate formulas below, the laser peak power 
$P_0 = \vec{B}_0^2 = \vec{E}_0^2$ 
and  its (angular) frequency width $\Delta \omega_0 = 2 \pi \Delta \nu_0$.
The width factor $ \gamma_p$ in the resonance formula is equal to
$ {\rm Max\,} ( \Delta \omega_0, $ inverse lifetime
of state $|p \rangle$ or $|e \rangle )$.

We fully exploit macro-coherent amplification mechanism 
\cite{renp overview} to enhance rates of weak processes.
The macro-coherence amplification has been
experimentally  verified in weak QED processes of two-photon
emission of para-hydrogen vibrational transitions \cite{psr exp}.
Compared with spontaneous emission rate, the amplification factor was
$\sim 10^{18}$ in these experiments.
Rates of macro-coherent amplification are in proportion to $n^2 V = n N$ where
$n$ is the number density of target atoms and $V$ is the volume
of target region with $N$ the number of total target atoms.
Macro-coherent rates are thus enhanced by $n$ in contrast
to spontaneous emission rate $\propto N$.
When macro-coherence works, both
the energy and the momentum conservation holds
(neglecting very small atomic recoil), to give rates given by
a phase-space integrated quantity of
\begin{eqnarray}
&&
|{\cal M}|^2 n^2 V (2\pi)^4 \delta^{(4)} ( p_e - p_g - k_0)
\,.
\end{eqnarray}

The phase-space integrated RENP rates are given by
\begin{eqnarray}
&&
\Gamma_{{\rm RENP 1}} =
\frac{G_F^2}{ 24 \pi} \frac{P_0}{\Delta \nu_0 } 
\frac{\gamma_{pg} }{\epsilon_{pg}^3 \gamma_p } |\rho_{eg}|^2  |\vec{S}_{ep} |^2%S(S+1)
n^2 V {\cal F}
\,, \hspace{0.5cm}
{\cal F} = \sum_{ij} {\cal F}_{ij} \Theta \left( {\cal M}^2  - (m_i + m_j)^2
\right)\,,
\label {renp rate}
\\ &&
4 \pi  {\cal F}_{ij} 
= 
\left\{
\left(1 - \frac{ (m_i + m_j)^2}{  {\cal M}^2} \right)
\left(1 - \frac{ (m_i - m_j)^2}{  {\cal M}^2} \right)
\right\}^{1/2} 
\left[
| b_{ij} |^2 \left({\cal M}^2 - m_i^2 - m_j^2  \right)
- 2 \delta_M  \Re( b_{ij}^2 )\,m_i m_j \right]
\,,
\nonumber \\ &&
\label {2nu integral pc}
\\ &&
 {\cal M}^2 = (\epsilon_{eg} - \omega_0 )^2 - (\vec{p}_{eg} - \vec{k}_0)^2
= (1-r^2 )\epsilon_{eg}^2 - 2 \epsilon_{eg} \epsilon_{pg} \left( 1 - r \cos \theta
\right) \equiv  {\cal M}^2(\theta)
\,,
\label {mass sq f}
\end{eqnarray}
for parity conserving part, and
\begin{eqnarray}
&&
\Gamma_{{\rm RENP 2}} =
\frac{G_F^2}{ 12 \pi} \frac{P_0}{\Delta \nu_0 } 
\frac{\gamma_{pg} }{\epsilon_{pg}^3 \gamma_p } |\rho_{eg}|^2  
\vec{S}_{ep}\cdot \frac{\vec{p}_{ep}}{m_e} %S(S+1)
n^2 V {\cal G}
\,, \hspace{0.5cm}
{\cal G} = \sum_{ij} {\cal G}_{ij} \Theta \left( {\cal M}^2  - (m_i + m_j)^2
\right)\,,
\label {renp rate 2}
\\ &&
\hspace*{-1cm}
4 \pi  {\cal G}_{ij} 
= 
\left\{
\left(1 - \frac{ (m_i + m_j)^2}{  {\cal M}^2} \right)
\left(1 - \frac{ (m_i - m_j)^2}{  {\cal M}^2} \right)
\right\}^{1/2} 
\left[
\Re(c_{ij}^* b_{ij} ) \left({\cal M}^2 - m_i^2 - m_j^2  \right)
- 2 \delta_M  \Re(c_{ij} b_{ij} )\,m_i m_j \right]
\,,
\nonumber \\ &&
\label {2nu integral pv}
\end{eqnarray}
for parity violating part.
Majorana/Dirac difference appears in terms $\propto \delta_M$:
$\delta_M = 1$ for Majorana neutrino due
to the effect of anti-symmetrized wave functions of
two identical fermions \cite{my-07},
 and = 0 for Dirac neutrino.
We neglected small contribution in $\Gamma_{{\rm RENP 1}} $ 
from squared vector current $\propto (\vec{p}_{pe}/m_e )^2$.

The mass squared function $ {\cal M}^2(\theta) $ 
is invariant mass squared given to
neutrino pairs and a function of measurable quantity $\theta$.
The same function determines six neutrino-pair emission thresholds 
at different directions $\theta$ given by $ {\cal M}^2(\theta) = (m_i + m_j)^2\,, i,j = 1,2,3$.
It also  plays important roles in QED background rejection, as discussed
below.
The initial spatial phase vector $\vec{p}_{eg} $ given at excitation
 was parametrized by $|\vec{p}_{eg}| = r \epsilon_{eg}$
with its direction $\theta $ away from the RENP trigger beam.

We  adopt for excitation scheme
two-photon  cascade process using two pulsed lasers:
$\gamma_1 + | g\rangle^{\pm} \rightarrow | q \rangle^{\mp} $ and 
$\gamma_2 + | q\rangle^{\mp} \rightarrow | e \rangle^{\pm} $.
This way two relevant states, $|g\rangle $ and $|e \rangle $,
have the same time reversal quantum numbers, T $= \pm$,
since single photon process is governed by time reversal
odd operator whether it is of magnetic dipole or electric
dipole.
Counter-propagating two-photon cascade excitation requires frequencies of
two lasers to satisfy
$\omega_1 + \omega_2 = \epsilon_{eg}\,,  
\omega_1 - \omega_2 = r \epsilon_{eg}$.
A maximal excitation rate is provided when two frequencies $\omega_i$ are matched to
level spacings, namely resonant excitation.
The neutrino-pair emission operator is also T-odd.
T-even  two-photon excitation here
helps to reject QED background processes, as discussed
later.

The quantity $\rho_{eg}$ (called coherence in the optics literature)
that appears in rate formulas, eq.(\ref {renp rate}) and  eq.(\ref {renp rate 2}),
is generated at excitation to $| e\rangle $, its maximum value being 1/2.
It is in general time dependent, too.
Estimate of this quantity requires detailed simulations
of Maxwell-Bloch equations,
a set of non-linear partial differential equations that
deal with ion state wave functions and propagating electromagnetic fields
in a simplified spacetime of one space and one time dimensions \cite{renp overview}.
In a off-resonance Raman excitation simulations
suggest that this quantity is of order $10^{-3}$ or less.
We expect a larger coherence of order $10^{-1} \sim 10^{-2}$
in the on-resonance cascade excitation adopted here.

Magnetization is a macroscopic quantity given by
\begin{eqnarray}
&&
2 \mu_{{\rm eff}} n \approx 6.6\, {\rm G} \frac{n}{ 10^{18}{\rm cm}^{-3}} 
\; ({\rm for \;  Er}^{3+})
\,,
\end{eqnarray}
using $\mu_{{\rm eff}} = g \sqrt{J(J+1)} \approx 9.5 \mu_B $ for Er$^{3+}$ in its ground state
as an illustration.
The generated magnetization by RENP is this value times
interference rate, $\Gamma_{{\rm RENP 2}} $ of eq.(\ref{renp rate 2}).

Both rates $\propto {\cal F}$ and magnetization $\propto {\cal G}$ are functions of
the mass squared function $ {\cal M}^2(\theta) $.
In measuring the angular distribution of signals
one may use a single trigger beam at different directions of irradiation
to determine neutrino-pair thresholds.
Six thresholds have different weights, which are  calculable from neutrino oscillation data
\cite{pdg} and
listed in the following table, assuming CP conservation case described by
real number $b_{ij}, c_{ij}$.

\vspace{0.5cm}
\hspace*{2cm}
\begin{tabular}{|c|c|c|c|c|c|c|} \hline 
&(11) & (12) & (22) & (13) & (23) & (33)\\ \hline \hline
{\rm rate} & 0.0311 & 0.405 & 0.0401  & 0.0325 & 0.0144 & 0.227 \\ \hline
{\rm magnetization} & 0.508 &0.405 & 0.517 & 0.0325 & 0.0144 & 0.704
 \\ \hline
%& & &  \\ \hline
\end{tabular}
\vspace{0.5cm}

Spin factors are calculated as follows.
We consider unpolarized targets, hence average over initial 
$|e \rangle$
magnetic quantum numbers and sum over final $|p \rangle$
 magnetic quantum numbers.
Using Wigner-Eckart theorem applied to manifolds of the same $J$ value
(in actual application $J=11/2$),
one may relate $\vec{S}_{ep}^2$ and 
$\vec{S}_{ep}\cdot \vec{p}_{ep}/m_e$
to electric and magnetic dipole transition rates.
Summation and average over magnetic $J$ quantum numbers 
of two state gives 
$( \vec{S}_{ep}\cdot \vec{p}_{ep}/m_e)^2 =  \vec{S}_{ep}^2 ( \vec{p}_{ep}/m_e)^2$.
Relation of spin and velocity matrix elements to transition rates is
$\gamma^E_{ep} = e^2 ( \vec{p}_{ep}/m_e)^2 \Delta_{ep}/3\pi$ 
and $\gamma^M_{ep} = \vec{S}_{ep}^2 \mu_{ep}^2\Delta_{ep}^3/3\pi$ 
where $\Delta_{ep}$ is
the energy difference of two states.
Hence $ ( \vec{p}_{ep}/m_e)^2/\vec{S}_{ep}^2 
= e^2/(\mu_{ep}^2\Delta_{ep}^2 ) = (2m_e/g \Delta_{ep})^2 $
with the Lande g-factor $g=2$ of electron spin.
This gives
\begin{eqnarray}
&&
\frac{|\vec{v}_{ep}| }{ | \vec{S}_{ep} |} 
= \frac{  \Delta_{ep} } { m_e} \sqrt{ \frac{ \gamma^{ E}_{ep} }{ \gamma^{M}_{ep} }} \equiv (\frac{v}{S})_{ep}
\,.
\end{eqnarray}
We illustrate an example of $v/S$ calculations
for Er$^{3+}$ in Appendix.
For unpolarized targets $\vec{S}_{ep}^2$ is of order unity, close to
$(2 J_p +1)/(2J_e+1) $.

To obtain a large PV/PC ratio it is desirable to use 
a large electric/magnetic transition ratio,
since the ratio $ \frac{\Delta_{pe}} {m_e}  \approx 10^{-6} $ is small.
In lanthanoid ions doped in crystals transitions among low lying levels are predominantly magnetic,
and it is desirable to have a large forced electric dipole transition.
For trivalent Kramers ions such as Er$^{3+}$
 placed at a less symmetric cite (not inversion center for instance) are necessary.
Even in cubic crystals trivalent lanthanoid ions often substitute 
at sites of less symmetry, C$_2$ site instead of C$_{3i}$ IC
(Inversion Center).
In this respect trivalent Er ion doped in host crystal such as YSO
may give larger magnetization than the example here.

\section
{\bf Dominant QED background}

It is easy to understand that both of macro-coherently amplified QED events,
$| e\rangle^{\pm} \rightarrow | g\rangle^{\pm} + \gamma \gamma $
and $| e\rangle^{\pm} \rightarrow | g\rangle^{\mp} + \gamma $ do not exist
by kinematic choice of a positive mass squared function 
$ {\cal M}^2(\theta)$,
which constrains trigger directions $\theta$ for a specified chosen parameter $r$.
Major backgrounds appear to be
macro-coherently amplified 
$| e\rangle^{\pm} \rightarrow | g\rangle^{\mp} + \gamma \gamma \gamma$ 
(called McQ3) and
spontaneous emission of a single photon.
McQ3 events are however characterized by a time reversal odd transition,
which is distinguished from time reversal even RENP process,
$| e\rangle^{\pm} \rightarrow | g\rangle^{\pm} + \gamma + \nu\bar{\nu}  $.
This way one can kill time reversal odd transition McQ3 and
spontaneous single photon emission, 
leaving McQ4 
$| e\rangle^{\pm} \rightarrow | g\rangle^{\pm} + 
\gamma \gamma\gamma \gamma$
the major QED background.

McQ4 do not contribute to magnetization.
The only thing one has to worry about is depletion of prepared state 
$| e\rangle^{\pm} $.
A crude estimate of McQ4 event rate shows that it is
a few to several orders of magnitudes larger than RENP rate.
With extra emitted photon detection,
this remaining background is presumably controllable even in rate measurements.

\section
{\bf Example of Er$^{3+}$ RENP scheme}

Trivalent Er ion has a rich J-manifold structure,
as illustrated in Appendix. 
The ion is at inversion center of this crystal, but
J-manifold structure is insensitive to crystal environments.
We consider the following de-excitation path of time
reversal degeneracy:
\begin{eqnarray}
&&
|e \rangle =^4{\rm H}_{11/2}^{\pm}(2.3946)  \rightarrow 
^4{\rm I}_{11/2}^{\mp}
|g \rangle =^4{\rm I}_{15/2}^{\pm} (0)
\,,
\label {higher level 2}
\end{eqnarray}
and  cascade excitation of frequencies,
\begin{eqnarray}
&&
\omega_1 =1.5894 {\rm eV}(= \epsilon(^4{\rm H}_{11/2} - ^4{\rm I}_{13/2})\,)
\,, \hspace{0.5cm}
\omega_2 =0.805   {\rm eV} (= \epsilon(^4{\rm I}_{13/2} - ^4{\rm I}_{15/2})\,)
\,, \hspace{0.5cm}
r = 0.3276
\,.
\end{eqnarray}
J-manifolds are split by crystal field, 
giving Stark levels having two-fold Kramers 
degeneracy.
We ignore effects of different Stark levels for simplicity.
In order to maintain the doubled rate of time reversal degeneracy,
it is necessary to shield earth magnetic field.

Using  Er$^{3+}$ data given in Appendix, we estimate
RENP rate $\Gamma$ and generated magnetization $M$:
\begin{eqnarray}
&&
\Gamma = 1.3 \times 10^{-3} {\rm sec}^{-1} \frac{{\cal F} }{{\rm eV}^2 } |\rho_{eg}|^2
|\vec{S}_{ep} |^2
( \frac{ n}{ 10^{18}{\rm cm}^{-3}})^2 \frac{V }{10^{-2}{\rm cm}^3 } 
\frac{P_0 }{{\rm GW cm}^{-2} } \frac{100 {\rm MHz} }{ \Delta \nu_0}
\,,
\label {rate number}
\\ &&
M = 7.3 \times 10^{-9} {\rm G\, sec}^{-1} 
\frac{{\cal G} }{{\rm eV}^2 } |\rho_{eg}|^2 |\vec{S}_{ep} |^2 
( \frac{ n}{ 10^{18}{\rm cm}^{-3}})^3 \frac{V }{10^{-2}{\rm cm}^3 } 
\frac{P_0 }{{\rm GW cm}^{-2} } \frac{100 {\rm MHz} }{ \Delta \nu_0}
\,.
\label {mag number}
\end{eqnarray}
For this estimate we  used numbers appropriate for pulsed trigger laser.
For (continuous wave) CW operation $P_0 = 1 {\rm kW cm}^{-2}\,,
\Delta \nu_0 = 1 {\rm kHz} $, hence 1/10 reductions
of rate and magnetization are more appropriate for CW.

Magnetization calculated this way is measurable, being 
comparable to, or  above,
the SQUID sensitivity level 
used in fundamental physics experiments such as in axion force experiments,
\cite{quax-gpgs}, \cite{axion force}, \cite{moody-wilczek} and axion haloscope
experiment \cite{axion haloscope}, \cite{sikivie}.

\begin{figure*}[htbp]
 \begin{center}
 \epsfxsize=0.8\textwidth
 \centerline{\includegraphics[width=7cm,keepaspectratio]{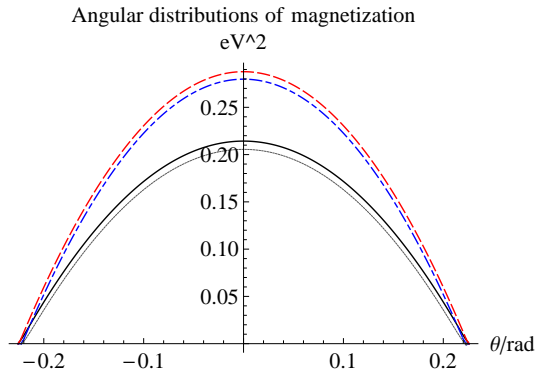}} \hspace*{\fill}
   \caption{ 
Angular $\theta$ distribution of magnetization
 given by $4\pi {\cal G} ( \theta) $ for $r= 0.3276$:
NH (Normal Hierarchy) 
Majorana neutrinos of smallest mass 5 meV in solid black, 50 meV in dotted black,
NH Dirac 5 meV in dashed red, and 50 meV in dash-dotted blue.
All CP violating phases are assumed to vanish for simplicity.
Absolute values of magnetization are obtained by multiplying
numbers here with $7.3 \times 10^{-9}$ G sec$^{-1}$ of eq.(\ref{mag number}).
}
   \label {angular dist 3}
 \end{center} 
\end{figure*}

\begin{figure*}[htbp]
 \begin{center}
 \epsfxsize=0.8\textwidth
 \centerline{\includegraphics[width=7cm,keepaspectratio]{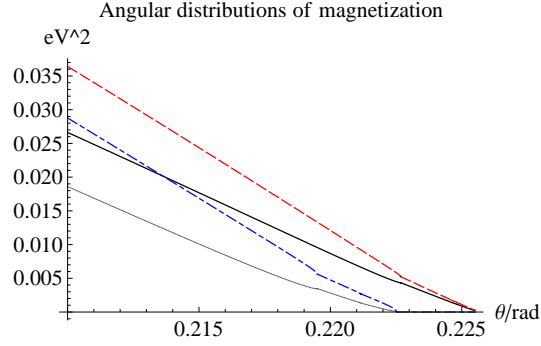}} \hspace*{\fill}
   \caption{ 
Enlarged end point region corresponding to Fig(\ref{angular dist 3}).
}
   \label {angular dist 4}
 \end{center} 
\end{figure*}

\begin{figure*}[htbp]
 \begin{center}
 \epsfxsize=0.8\textwidth
 \centerline{\includegraphics[width=7cm,keepaspectratio]{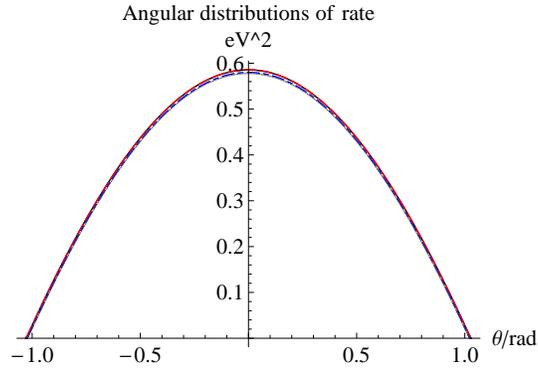}} \hspace*{\fill}
   \caption{ 
Angular $\theta$ distribution of rate given by $4\pi {\cal F} ( \theta) $ for $r= 0.3276$:
NH Majorana of smallest mass 5 meV in solid black, 50 meV in dotted black,
NH Dirac 5 meV in dashed red, and 50 meV in dash-dotted blue.
All CP violating phases are assumed to vanish for simplicity.
Absolute values of rate are obtained by multiplying
numbers here with $1.3 \times 10^{-3}$ sec$^{-1}$ of eq.(\ref{rate number}).
}
   \label {rate angular dist}
 \end{center} 
\end{figure*}

We show magnetization curves, its angular $\theta$ distributions
at a specific excitation parameter $r$,
in Fig(\ref {angular dist 3}) and Fig(\ref {angular dist 4}).
For comparison rate angular distribution is shown in  Fig(\ref {rate angular dist}).
Magnetization depends on $b_{ij} \times c_{ij}$ above $(ij)$ pair thresholds, while
rate depends on $b_{ij}^2$.
This difference, as numerically shown in the table,
explains a high sensitivity of magnetization to the Majorana/Dirac distinction.
On the other hand, it is difficult to distinguish mass types from rate, as seen in Fig(\ref{rate angular dist}).
Absolute mass determination is harder, and one needs a high statistics data
near the end point of angular distribution, as shown in Fig(\ref {angular dist 4}).

\section
{\bf Towards Er$^{3+}$ experimental design}

There are many works to be done before we prepare technical design report of
actual experiments.

First,  results in the present work are incomplete in several places:

(1) Lack of theoretical calculations or experimental results
of crystal field effect in a variety of host crystals, namely, Stark levels and forced electric dipole
transition rates along with magnetic transition among low lying levels
for trivalent Er ions.

(2) Simulation estimate of the coherence $\rho_{eg}(t)$ assuming
excitation scheme and laser specification.

(3) Simulation of McQ4 QED backgrounds
in order to identify and isolate background events.

(4) Design of SQUID detection system.

Experimental R and D studies are even more important.
One may list some items:

(5) Optical measurements of 4f$^{11}$ manifolds in candidate host crystals.

(6) Study of relaxation processes, in particular, phonon related
relaxation at low temperatures.

\section
{\bf Appendix:  Er$^{3+}$ data and estimated parity violating component}

We consider a special host crystal,
Er$^{3+}$:Cs$_2$NaYF$_6$ or :Y$_2$O$_3$ for concreteness.
Trivalent Er ion substitutes inversion center of Y.
According to \cite{er3+}, calculated radiative decay rates of 10\% doped Er$^{3+}$ in host Cs$_2$NaYF$_6$ are

%\vspace{0.5cm}
\begin{eqnarray*}
\begin{array}{cccccc}
{\rm initial }  & {\rm final }  & {\rm energy/eV } & {\rm rate/sec}^{-1}  & {\rm radiative\; life/msec} & v/S (10^{-6} )\\
^4{\rm I}_{13/2} \rightarrow  &  ^4{\rm I}_{15/2} & 0.805 & 24.82^{{\rm MD}} + 2.46^{{\rm ED}} & 36.7 & 0.25 \\
^4{\rm I}_{11/2} \rightarrow & ^4{\rm I}_{15/2} & 1.286 & 4.13^{{\rm ED}} & 113.4 &\\
 & ^4{\rm I}_{13/2} & 0. 481 & 4.26^{{\rm MD}} + 0.42^{{\rm ED}}& & 0.15 \\
^4{\rm I}_{9/2} \rightarrow & ^4{\rm I}_{15/2} & 1.563 & 9.46^{{\rm ED}} & 73.3 &\\
& ^4{\rm I}_{13/2}  & 0.7575 & 3.48^{{\rm ED}} & & \\
& ^4{\rm I}_{11/2}  & 0.277 &  0.67^{{\rm MD}} + 0.02^{{\rm ED}}  & & 0.047 \\
^4{\rm F}_{9/2} \rightarrow & ^4{\rm I}_{15/2} & 1.9014 &58.91^{{\rm ED}} &14.6 & \\
& ^4{\rm I}_{13/2}  & 1.0964 & 3.5^{{\rm ED}}  & & \\
& ^4{\rm I}_{11/2}  & 0.619 &  3.49^{{\rm MD}} +1.06^{{\rm ED}}     & &  0.33\\
& ^4{\rm I}_{9/2}  & 0.339 &   1.34^{{\rm MD}} +0.18^{{\rm ED}}   & & 0.12\\
^4{\rm S}_{3/2} \rightarrow  &  ^4{\rm I}_{15/2} & 2.2645 {\rm eV} & 4.93^{{\rm ED}}  & 131.6 & \\
 & ^4{\rm I}_{13/2}  &  1.4593{\rm eV} & 2.02^{{\rm ED}}  & & \\
 & ^4{\rm I}_{11/2}  &  0.9785{\rm eV} &  0.18^{{\rm ED}}  & &\\
 &^4{\rm I}_{9/2} &  0.7017{\rm eV} &  0.48^{{\rm ED}}  & &\\
^4{\rm H}_{11/2} \rightarrow  &  ^4{\rm I}_{15/2} & 2.3946 {\rm eV} &518.02^{{\rm ED}}  & 1.7 & \\
 & ^4{\rm I}_{13/2}  &  1.5894{\rm eV} &  49.43^{{\rm MD}}+7.91^{{\rm ED}}   & & 0.62 \\
 & ^4{\rm I}_{11/2}  &  1.1091{\rm eV} &   5.58^{{\rm MD}} +4.73^{{\rm ED}} & & 1.0\\
 &^4{\rm I}_{9/2} &  0.8318{\rm eV} &   0.49^{{\rm MD}} + 5.52^{{\rm ED}}& & 2.7
\end{array}
\end{eqnarray*}

Low lying levels of Er$^{3+}$ have configuration 4f$^{11}$ and its ground state  $^4{\rm I}_{15/2}$
has the effective magnetic moment related to the Lande factor $\mu_{{\rm eff}} = g \sqrt{J(J+1)} = 9.5$ (experimental value).
Ratio of parity violating to conserving amplitude given by $v/S$ was
estimated by using the formula in the text.
Low values of $v/S$ here are presumably related to Er$^{3+}$ site at inversion center (IC) of host crystals.
Without IC $v/S$'s  are expected to be larger.

\vspace{1cm}
 {\bf Acknowledgements}

This research was partially
 supported by Grant-in-Aid 17H02895 %(MY)  
from the
 Ministry of Education, Culture, Sports, Science, and Technology.

\end{document}